\documentclass[apj]{emulateapj}

\usepackage{graphicx,amsmath,enumerate}
\usepackage{natbib,times}
\usepackage{verbatim, color, soul}

\newcommand{\be}{\begin{equation}}
\newcommand{\ee}{\end{equation}}

\newcommand{\kms}{\,{\rm {km\, s^{-1}}}}
\newcommand{\msun}{{$M_{\odot}$}}

\newcommand{\ergs}{erg~s$^{-1}$}
\newcommand{\esc}{erg~s$^{-1}$~cm$^{-2}$}

\newcommand{\gtsima}{$\; \buildrel > \over \sim \;$}
\newcommand{\ltsima}{$\; \buildrel < \over \sim \;$}
\newcommand{\prosima}{$\; \buildrel \propto \over \sim \;$}
\newcommand{\gsim}{\lower.5ex\hbox{\gtsima}}
\newcommand{\lsim}{\lower.5ex\hbox{\ltsima}}
\newcommand{\simgt}{\lower.5ex\hbox{\gtsima}}
\newcommand{\simlt}{\lower.5ex\hbox{\ltsima}}
\newcommand{\simpr}{\lower.5ex\hbox{\prosima}}

\newcommand{\etal}{{et al.~}}

\begin{document}

\title{A $\sim50,000~M_{\odot}$ black hole in the nucleus of RGG 118}

\author{Vivienne F. Baldassare\altaffilmark{1}, Amy E. Reines\altaffilmark{1,2}, Elena Gallo\altaffilmark{1}, Jenny E. Greene\altaffilmark{3}}
\altaffiltext{1}{Department of Astronomy, University of Michigan, Ann Arbor, MI 48109}
\altaffiltext{2}{Hubble Fellow}
\altaffiltext{3}{Department of Astrophysical Sciences, Princeton University, Princeton, NJ 08544}

\begin{abstract}
Scaling relations between black hole (BH) masses and their host galaxy properties have been studied extensively over the last two decades, and point towards co-evolution of central massive BHs and their hosts. However, these relations remain poorly constrained for BH masses below $\sim10^{6}$ \msun. Here we present optical and X-ray observations of the dwarf galaxy RGG 118 taken with the Magellan Echellette Spectrograph on the 6.5m Clay Telescope and \textit{Chandra} X-ray Observatory. Based on Sloan Digital Sky Survey spectroscopy, RGG 118 was identified as possessing narrow emission line ratios indicative of photoionization partly due to an active galactic nucleus. Our higher resolution spectroscopy clearly reveals broad H$\alpha$ emission in the spectrum of RGG 118. Using virial BH mass estimate techniques, we calculate a BH mass of $\sim50,000$ \msun. We detect a nuclear X-ray point source in RGG 118, suggesting a total accretion powered luminosity of $L=4\times10^{40}~{\rm erg~s^{-1}}$, and an Eddington fraction of $\sim1$ per cent. The BH in RGG 118 is the smallest ever reported in a galaxy nucleus and we find that it lies on the extrapolation of the $M_{\rm BH}-\sigma_{\ast}$ relation to the lowest masses yet. 
\end{abstract}

\section{Introduction}

It is now established that most, if not all, massive galaxies (e.g. $M_{\ast}\gtrsim10^{10}$ \msun) host massive black holes (BHs) at their centers. Moreover, the BH masses correlate with properties of the host galaxy, such as bulge stellar velocity dispersion, bulge luminosity, and bulge mass, pointing towards co-evolution of galaxies and their BHs \citep{2000ApJ...539L...9F, 2000ApJ...539L..13G, Marconi:2003fk, 2009ApJ...698..198G, Kormendy:2013ve, 2013ApJ...764..184M}. However, these scaling relations are poorly constrained for low BH/galaxy masses ($M_{\rm BH}\lesssim10^{6}$ \msun). It is presently unknown what fraction of low-mass galaxies ($M_{\ast}\lesssim10^{10}$ \msun) contain BHs, and BH mass measurements in this regime are relatively scarce. 

The low-mass end of scaling relations is of particular importance for constraining models of high-redshift BH seed formation. \cite{2009MNRAS.400.1911V} found that for BH masses $\lesssim10^{6}$ \msun, the slope and scatter of $M_{\rm BH}-\sigma_{\ast}$ vary depending on whether BH seeds were massive ($\sim10^{4-5}$ \msun, e.g. from direct collapse) or smaller Population III remnants ($\sim100$ \msun). Additionally, the fraction of low-mass galaxies hosting BHs is itself important for models of BH seed formation (e.g. \citealt{2008MNRAS.383.1079V, 2015ApJ...799...98M}). 

BH mass measurements for dwarf galaxies remain difficult to achieve and are often highly uncertain. Dynamical mass measurements for low-mass BHs are limited by the resolving power of our telescopes; even with the \textit{Hubble Space Telescope}, we can only resolve the sphere of influence of BHs with $M_{\rm BH}\sim10^{5}$ \msun~for the Local Group. Active galactic nuclei (AGN) offer an opportunity to measure BH masses in systems beyond the Local Group; for AGN in dwarf galaxies, we can estimate masses based on broad emission lines if present (e.g. \citealt{2005ApJ...630..122G, Reines:2013fj}), or from the fundamental plane of BH activity \citep{2003MNRAS.345.1057M, Reines:2011fr, 2014ApJ...787L..30R}. These methods rely on correlations between various observables and BH mass, and thus carry larger uncertainties.

In the last few years, the number of known dwarf galaxy--AGN systems has moved beyond a few key examples such as NGC~4395 \citep{2003ApJ...588L..13F}, POX~52 \citep{2004ApJ...607...90B}, and Henize 2-10 \citep{Reines:2011fr, 2012ApJ...750L..24R} to increasingly larger samples. The first systematic searches for low-mass BHs were done by \cite{2004ApJ...610..722G, 2007ApJ...670...92G}, which searched the Sloan Digital Sky Survey (SDSS) for galaxies with broad H$\alpha$ emission indicating AGN with masses below $2\times10^{6}M_{\odot}$. Though these surveys identified $\sim200$ such BHs, the host galaxies had stellar masses larger than typical dwarf galaxies (\citealt{Greene:2008qy}). More recently, \cite{Reines:2013fj} increased the number of known AGN in dwarf galaxies by an order of magnitude by searching for narrow and broad emission line AGN signatures in a sample of nearby (z $\lesssim0.055$) dwarf galaxies ($M_{\ast}\lesssim3\times10^{9}$ \msun) in the SDSS (see also \citealt{2014AJ....148..136M}).

Here we present optical spectroscopic and X-ray observations of SDSS J1523+1145 (object 118 in \citealt{Reines:2013fj}; hereafter referred to as RGG 118), a dwarf disk galaxy with a stellar mass of $\sim2.5\times10^{9}$ \msun~at a redshift of z = 0.0243. First identified as having narrow line ratios indicative of AGN activity based on SDSS spectroscopy, our new higher resolution spectroscopy reveals broad H$\alpha$ emission indicative of a $\sim50,000$ \msun\ BH.

\section{Observations and Data Analysis}

\subsection{Spectroscopy}
\subsubsection{SDSS}

\begin{figure*}
\centering
\includegraphics[scale=0.9]{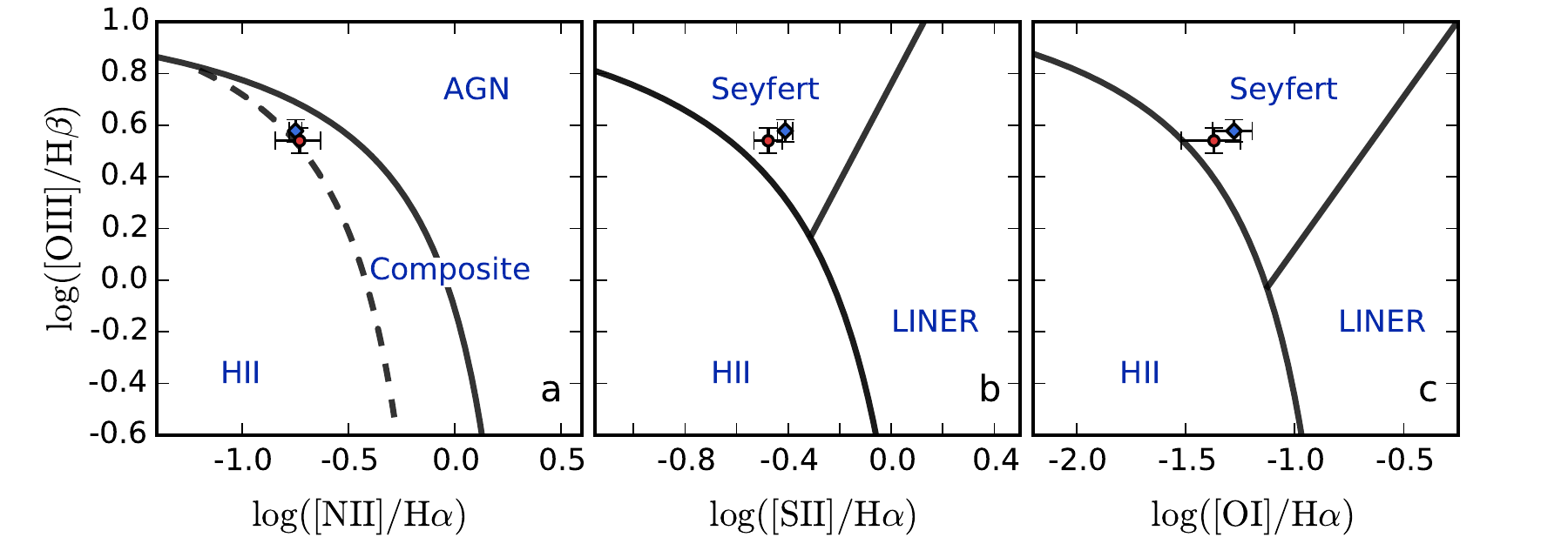}
\caption{Diagnostic diagrams which characterize the photoionizing continuum in RGG 118 \citep{1981PASP...93....5B,1987ApJS...63..295V, 2006MNRAS.372..961K}. All diagrams plot $\log([{\rm OIII]/H}\beta)$ on the y-axis. On the x-axis, ratios of $[\rm{NII}]\lambda6583$, $[\rm{SII}]\lambda6716,6731$, and $[\rm{OI}]\lambda6300$ to H$\alpha$ are shown from left to right. Blue diamonds represent values from SDSS data (Reines \etal 2013) while red circles represent MagE data. All three diagrams indicate that the photoionization in RGG 118 is at least partly due to an AGN.}
\label{diag}
\end{figure*}

\cite{Reines:2013fj} used the SDSS and NASA-Sloan Atlas to search for dwarf galaxies with optical spectroscopic signatures of accretion onto a massive BH. They modeled the spectra of $\sim$25,000 nearby (z $\lesssim0.055$) dwarf ($M_{\ast}\lesssim3\times10^{9}$ \msun) galaxies, and measured emission line fluxes. Using the narrow-line [OIII]/H$\beta$ versus [NII]/H$\alpha$ diagnostic diagram \citep{2006MNRAS.372..961K}, they identified 136 galaxies with emission line ratios consistent with (at least some) photoionization from an accreting BH. RGG 118 fell into the ``composite" region of the [OIII]/H$\beta$ versus [NII]/H$\alpha$ diagram, indicating photoionization from both an accreting BH and stellar processes (Figure~\ref{diag}). The SDSS spectrum of RGG 118 (14 May 2007) shows evidence for broad H$\alpha$ emission (see top right of Figure~\ref{magespec}), but it was not identified by Reines \etal (2013) because it falls below their broad H$\alpha$ detection limit of $\sim2\times10^{39}$ \ergs\ (at their sample median redshift of z$\sim$0.03). 

\subsubsection{MagE}

Spectra of RGG 118 were taken on 18 April 2013 with the 6.5m Magellan II telescope at Las Campanas Observatory, using the Magellan Echellette Spectrograph (MagE; \citealt{2008SPIE.7014E..54M}). MagE is a moderate resolution ($\lambda/\Delta\lambda = 4100$) echelle spectrograph, with wavelength coverage extending from 3000 to 10000$\rm \AA$ over 15 orders. Using the arc lamp exposures, we measure an instrumental dispersion $\sigma_{\rm instrument}=28~\rm{km}~\rm{s}^{-1}\pm5~\rm{km}~\rm{s}^{-1}$. Two 1200-second exposures of RGG 118 were taken using a 1\arcsec\ slit. Flat fielding, sky subtraction, extraction, and wavelength calibration were carried out using the mage\_reduce IDL pipeline written by George Becker. We used a boxcar extraction with an extraction width of 1\arcsec, in order to obtain the nuclear spectrum and reduce contamination from starlight.  

\begin{figure*}
\centering
\includegraphics[scale=0.72]{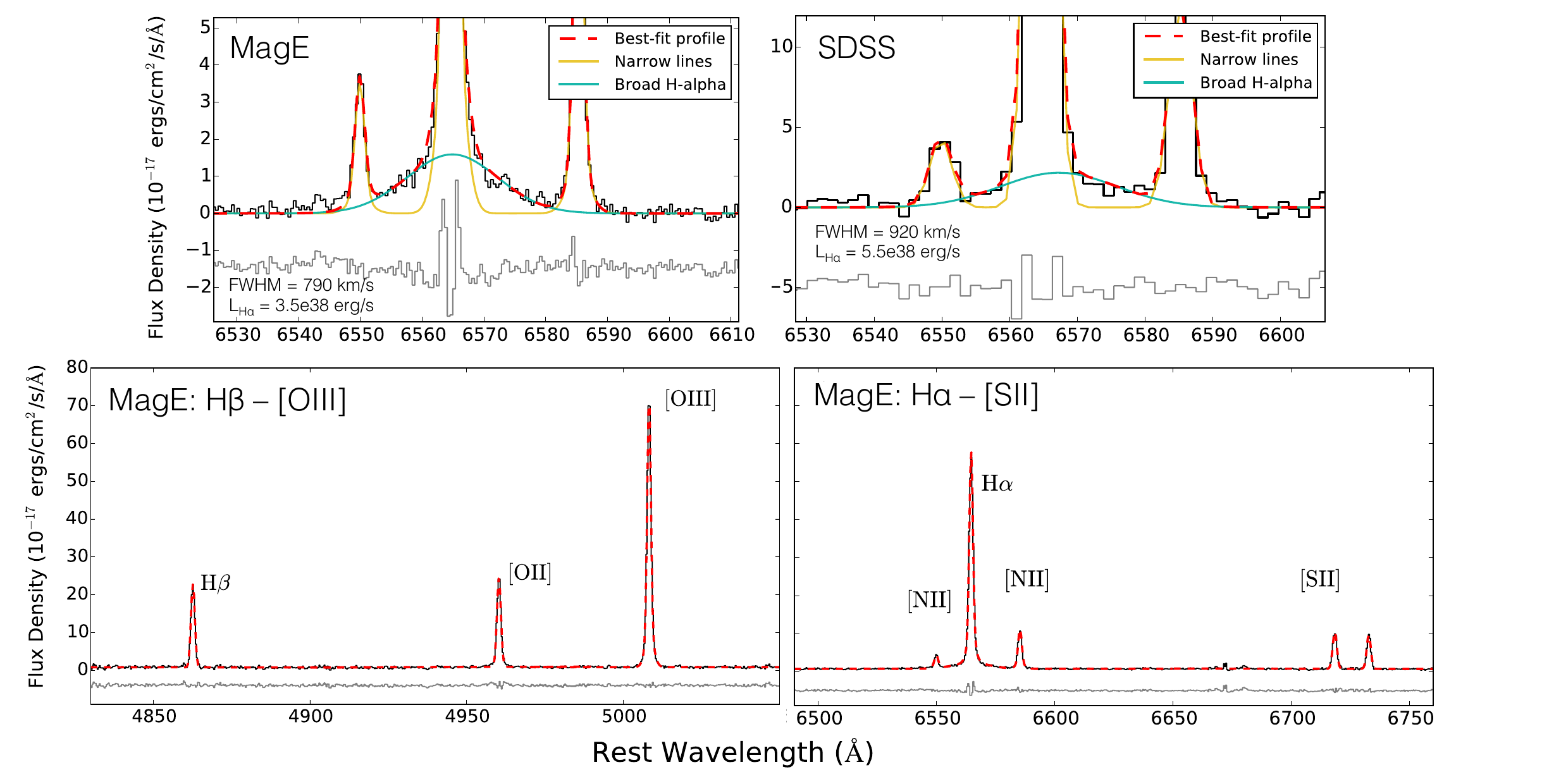}
\caption{In all panels, the black line is the observed spectrum, the red dashed line is the best-fit total profile, and the gray line is the residual between the observed spectrum and best fit, offset by an arbitrary amount.{\it Top}: MagE (left) and SDSS (right) spectra showing the H$\alpha$-[NII] complex. The yellow and teal solid lines represent the best-fit narrow and broad components, respectively.  The narrow line model used for the MagE fit shown here is based on the [OIII] line (Model C; see text). The SDSS best-fit parameters yield a BH mass of $\sim91,000$ \msun. The MagE H$\alpha$ FWHM and luminosity imply a BH mass of $\sim50,000$ \msun.  {\it Bottom}: Regions of the MagE spectrum of RGG 118 showing the emission lines most relevant to our analysis.}
\label{magespec}
\end{figure*}

We analyzed the spectrum of RGG 118 using custom emission line fitting software developed in Python and broadly following the procedure laid out in \cite{Reines:2013fj}. Our software uses non-linear least squares minimization to fit Gaussian profiles to relevant emission lines. 

Fitting the H$\alpha$-[NII] complex is of particular importance because the luminosity and FWHM of broad H$\alpha$ emission can be used to estimate BH mass \citep{2005ApJ...630..122G}. The general procedure is to obtain a model for an intrinsically narrow line, then use that model to fit H$\alpha$ $\lambda$6563, [NII] $\lambda$6548, and [NII] $\lambda$6583 simultaneously. We then add an additional component to model any broad excess and include this component in our model if its inclusion improves the reduced $\chi^{2}$ by at least 20\%. We then measure line widths and fluxes for the [NII] lines, and narrow and broad H$\alpha$. 

We did not detect stellar absorption lines in the RGG 118 spectrum, so we model the continuum in each order as a line and subtract it before modeling the emission lines. We use three different methods to model the narrow line emission and find that an additional broad component for H$\alpha$ is required in all three cases. We first use a single Gaussian narrow line model based on the profiles of the [SII] $\lambda\lambda$6713,6731 doublet (Model A). Next, we use a narrow line model based on the profile of [OIII] $\lambda$5007, which is also a forbidden transition. [OIII] $\lambda$5007 is significantly brighter than [SII], and can reveal the presence of wings \citep{2013MNRAS.433..622M, 2014MNRAS.442..784Z}. Indeed, we modeled [OIII] $\lambda$5007 in two different ways and found both required a narrow core and wing component. First, we use the [SII] profile to model the core of the [OIII] line and find a wing component with FWHM$\sim330 \kms$ (Model B). In the second [OIII] model we fit the core with no constraints from [SII]; this yields a wing of FWHM $\sim192 \kms$ (Model C). Finally, we use each best fitting [OIII] profile as a model for the narrow line emission in the H$\alpha$-[NII] complex. See Table~\ref{halphas} for broad H$\alpha$ FWHM and luminosities corresponding to each narrow line model. 

\begin{table}[h]
\caption{}
\begin{tabular}{c|cc|c}
Model & $L_{\rm H\alpha}$ (\ergs) & FWHM$_{\rm H\alpha}$ ($\kms$) & BH mass ($10^{4}$ \msun) \\
\hline
A & $4\times10^{38}$ & 543 & 2.7 \\
B & $3\times10^{38}$ & 855 & 6.2 \\
C & $3.5\times10^{38}$ & 790 & 5.4 \\
\end{tabular}
{\bf Notes.} Broad H$\alpha$ luminosities and FWHM for narrow line models A, B, and C. Also given are the corresponding BH masses for each model.
\label{halphas}
\end{table}

We also model the [OII] $\lambda$3727, H$\beta$, [OIII] $\lambda$4959 and [OI] $\lambda$6300 lines and place RGG 118 on narrow line diagnostic diagrams (Figure~\ref{diag}), and find the MagE narrow line ratios to be consistent with those found in the SDSS spectrum. A broad component was not found to be statistically justified for the H$\beta$ emission line of RGG 118 (to be expected given the low luminosity of broad H$\alpha$ and the typical H$\alpha$/H$\beta$ intensity ratio of $\sim$3). We allow up to two Gaussians for the oxygen lines, which sometimes can exhibit broadening from the narrow line region. Figure ~\ref{magespec} shows the fitted MagE spectrum, as well as a close-up of the H$\alpha$-[NII] complex for Model C. 

\subsection{Chandra X-ray Observatory}

\textit{Chandra} observed the field of RGG 118 with the Advanced CCD Imaging Spectrometer (ACIS) detector on 2014-12-26 UT 18:09:21, for a total of 19.8 ksec, with the target galaxy placed on the back-illuminated S3 chip of ACIS-S. 
Data were telemetered in very-faint mode and analysed with the \textit{Chandra} Interactive Analysis of Observations software package (CIAO, version 4.6). 
First, we improved the \textit{Chandra} astrometry by cross-matching the detected X-ray point sources on the S3 chip to the SDSS catalog (the initial X-ray source list was derived by running CIAO WAVDETECT on the pipeline event file, with the exclusion of a 2\arcsec\ circular aperture centered on the nominal position of RGG 118). This resulted in 4 matches to an SDSS optical counterpart with $r$-band magnitude brighter than 23 mag, yielding a final astrometric correction of $\sim$0\arcsec.7. 
Next, we reprocessed the data to generate a new event list, and searched for time intervals with anomalously high background rates (none were found). Further analysis was restricted to the energy interval 0.5-7 keV, where ACIS is best calibrated. 

\begin{figure*}
\centering
\includegraphics[scale=0.65]{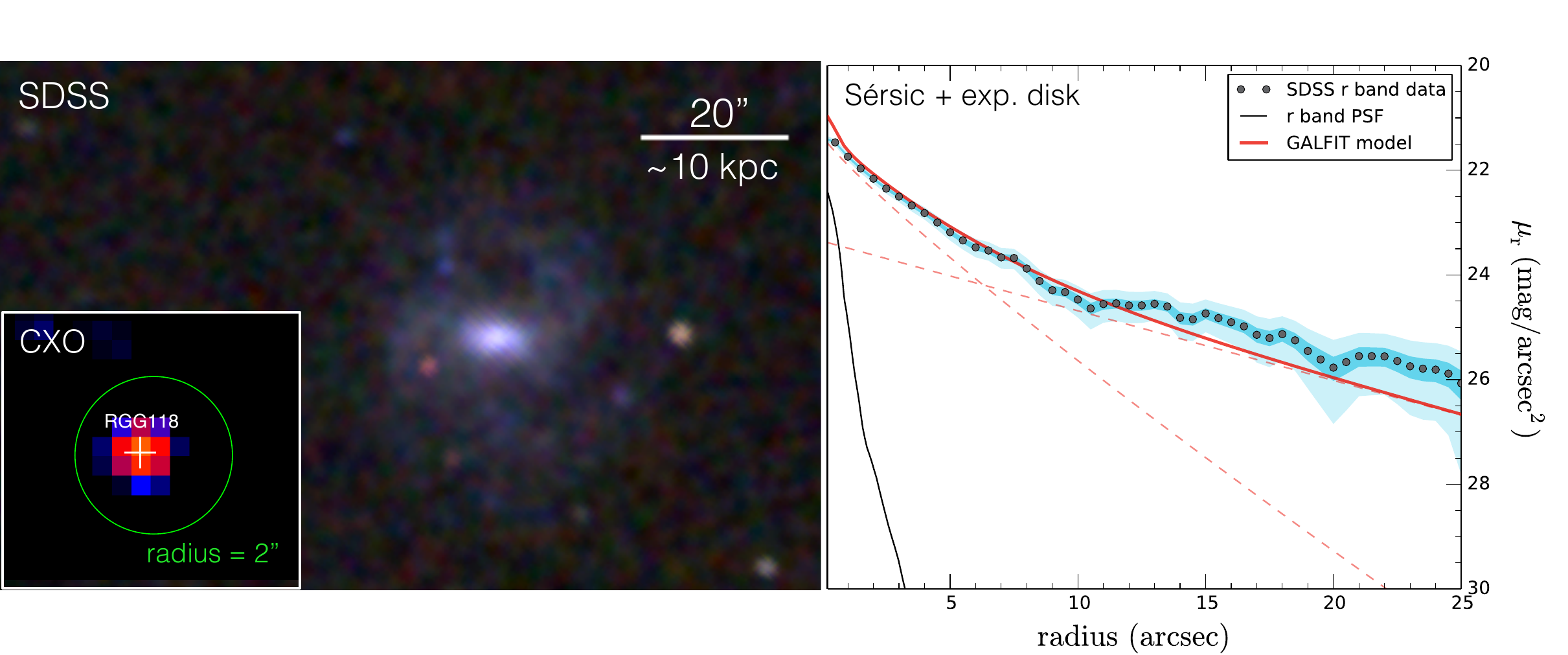}
\caption{{\it Left}: Smoothed SDSS \textit{gri} image of RGG 118 (filters colored blue, green, and red, respectively). Inset: Chandra ACIS-S image of the field of view of RGG 118. An X-ray source is clearly detected at a position consistent with the nominal SDSS position (indicated by a green circle with a radius of 2\arcsec). The image has been smoothed with Gaussian kernel of $\sigma=3$ pixels. {\it Right}: \textit{r} band surface brightness versus radius for RGG 118. Surface brightness profile measurements from IRAF's ELLIPSE package are shown as gray dots. The dark and light blue shading represents the 1$\sigma$ and 3$\sigma$ uncertainties, respectively, due to the sky background. The solid black lines plot the \textit{r}-band PSF. The solid red line shows the best-fit S\'{e}rsic+exponential disk profile as determined by GALFIT (dashed red lines are individual components). }
\label{sdss_cxo}
\end{figure*}

A point-like X-ray source was detected at a position consistent with the SDSS position of RGG 118 (inset of Figure~\ref{sdss_cxo}; the X-ray source coordinates are R.A.: 15:23:05.0, Dec: $+$11:45:53.18; SDSS coordinates are R.A.: 15:23:04.97, Dec: $+$11:45:53.6). 
For the aperture photometry, we made use of CIAO SCRFLUX, which adopts a Bayesian formalism to estimate the source net count rate and corresponding flux. Source parameters were extracted from a 2\arcsec\ radius circular region centered at the source peak, while a (source-free) annulus of inner and outer radius of 20 and 35\arcsec\ was chosen for the background.
The low number of X-ray counts did not allow for a proper spectral modeling. Assuming that the source spectrum is well represented by an absorbed power-law model with photon index $\Gamma= 1.7$ (typical of actively accreting black holes) and equivalent hydrogen column $N_{\rm H}=3 \times 10^{20}$ cm$^{-2}$, the measured net count rate (1.96$\pm$0.6 $\times 10^{-4}$ counts sec$^{-1}$ between 0.5-7 keV) corresponds to an unabsorbed flux $F_{\rm X} = 3.4 \times 10^{-15}$ \esc\ over the energy interval 2-10 keV (90\% confidence intervals give a range from $9.1\times10^{-16}$ \esc~ to $8.1\times10^{-15}$ \esc). For a distance of 100 Mpc, this gives a hard X-ray luminosity of $L_{\rm 2-10~keV} = 4\times10^{39}$\ergs\ (90\% confidence intervals give a range from $1.1\times10^{39}$\ergs~ to $9.6\times10^{39}$\ergs).

\subsection{SDSS Imaging}
Five band {\it ugriz} imaging was taken for the field containing RGG 118 on 12 May 2005 (Figure~\ref{sdss_cxo}). In order to analyze the morphology of the host galaxy, we carried out a 2-D image decomposition of the SDSS {\it r}-band image using GALFIT \citep{2002AJ....124..266P,  2010AJ....139.2097P}. Using a two-tailed F test, we compared a single S\'{e}rsic model to a S\'{e}rsic plus exponential disk model and find that the latter is preferred \citep{2011ApJS..196...11S} (the reduced $\chi^{2}$ for single S\'{e}rsic model is 9.4, and the reduced $\chi^{2}$ for the S\'{e}rsic plus exponential disk model is 7.2). We measure a bulge ellipticity of 0.63. Figure~\ref{sdss_cxo} shows the best-fit surface brightness profile model. %

We use the integrated \textit{g} and \textit{r} band magnitudes of each component to determine the {\it r}-band mass-to-light ratio \citep{2003ApJS..149..289B}, and calculate stellar masses of the inner and outer (exponential disk) components. Magnitudes were corrected for reddening using galactic dust extinction maps which indicate \textit{E(B-V)} = 0.032 for RGG 118 \citep{2011ApJ...737..103S}.
For the inner S\'{e}rsic component (consistent with a pseudobulge, as discussed in Section~3), we measure $g-r=0.49\pm0.19$, which gives $\log_{10}({\rm M/L})_{r} = 0.23$ and $M_{\ast,{\rm inner}}$ = $10^{8.8 \pm 0.2}M_{\odot}$. 
For the exponential disk component we measure $g-r=0.41\pm0.07$, which yields $\log_{10}({\rm M/L})_{r} = 0.15$ and $M_{\ast,{\rm disk}}$ = $10^{9.3 \pm 0.1}M_{\odot}$. Combined, the masses of the bulge and disk are consistent with the NASA-Sloan Atlas total stellar mass of $M_{\ast}$ = $10^{9.3} M_{\odot}$. We also obtain a bulge-to-disk mass ratio of $\sim0.3$. 

\section{Black hole mass and scaling relations}
 
Strong evidence for BH accretion in RGG 118 comes from narrow emission line ratios (Figure~\ref{diag}), broad H$\alpha$ emission, and a nuclear X-ray point source (see Section 2.2). Assuming the broad emission is indeed the signature of gas rotating around a nuclear BH,  we can estimate the BH mass using standard virial techniques which assume the gas in the broad line region surrounding the BH is in virial equilibrium \citep{2004ApJ...613..682P, 2010ApJ...716..993B}. The virial relationship gives $M_{\rm BH} \propto R_{\rm BLR}\Delta V^{2} / G$, where $R_{\rm BLR}$ is the radius of the broad line emitting region, and $\Delta V$ is a characteristic velocity of gas in the broad line region. We use the H$\alpha$ emission line to estimate $\Delta V$ \citep{2005ApJ...630..122G}, and $L_{\rm H\alpha}$ as a proxy for $R_{\rm BLR}$ \citep{2000ApJ...533..631K, 2004ApJ...613..682P, 2005ApJ...630..122G, 2009ApJ...705..199B, 2013ApJ...767..149B}. The final equation used to estimate BH mass is given in \cite{Reines:2013fj}; we also adopt a scale factor of $\epsilon = 1$.

Each of the empirical relations used for this BH estimation technique has an associated scatter which contributes towards an uncertainty in the BH mass. Additionally, the scatter in the $M_{\rm BH}-\sigma_{\ast}$ relation ($\gtrsim0.3$ dex; \citealt{2012ApJ...753..125S}) also contributes to the total uncertainty, since the dimensionless scale factor $\epsilon$ is determined by calibrating the ensemble of reverberation-mapped AGN against the local $M_{\rm BH}-\sigma_{\ast}$ relation for non-active galaxies. Summing in quadrature gives a total uncertainty of 0.42 dex (a factor of $\sim2.7$).

For our three different narrow line modeling techniques (see Section 2.1.2), we obtain BH mass estimates ranging from $2.7\times10^{4}-6.2\times10^{4}$ \msun\ (Table~\ref{halphas}) and thus adopt a nominal BH mass of $\sim5\times10^{4}$\msun. This is the smallest mass reported for a BH in a galaxy nucleus. 
Additionally, assuming a X-ray to bolometric correction of 10 \citep{2004MNRAS.351..169M}, we calculate a total accretion-powered luminosity $L = 4\times 10^{40}$ \ergs\ for the black hole in RGG118. This corresponds to an Eddington ratio of $\sim1$ per cent, similar to AGN in more massive systems.  

While the broad H$\alpha$ emission in typical AGN (i.e. FWHM $\gtrsim$ thousands of $\kms$) is due to gas dominated by the gravity of a BH, some dwarf AGN have broad H$\alpha$ line widths consistent with broadening from stellar processes. Thus, we have also considered the following alternate explanations for the broad H$\alpha$ in RGG 118.  
\begin{enumerate}[(i)]

\item ~Supernovae: The broad Balmer emission and luminosities of Type II supernovae can often cause their spectra to resemble those of AGN \citep{1989AJ.....97..726F}. However, a supernova is a transient event; the marginal detection of broad H$\alpha$ seen in the SDSS spectrum taken 6 years prior to the MagE spectrum makes this an unlikely source of the broad H$\alpha$. 

\item ~Wolf-Rayet stars: Outflows from Wolf-Rayet stars can produce broad spectral features. However, the spectra of galaxies with Wolf-Rayet stars typically feature a characteristic ``bump" at $\lambda$4650 -- 4690 \citep{2000ApJ...531..776G}; we do not observe this feature in either the SDSS or MagE spectra of RGG 118. 

\item ~Luminous blue variables: Outbursts from luminous blue variables can produce broad H$\alpha$ emission, but they are fainter than Type II supernovae \citep{2011MNRAS.415..773S}. Additionally, this phenomenon is transient and so is also ruled out by the persistent broad H$\alpha$ emission of RGG 118.
\end{enumerate}

We do not consider the above to be likely explanations for the detected broad H$\alpha$ emission. Additional spectroscopic observations can further rule out the possibility of a stellar origin for the broad H$\alpha$ emission. %

\begin{figure*}[t]
\centering
\includegraphics[scale=0.63]{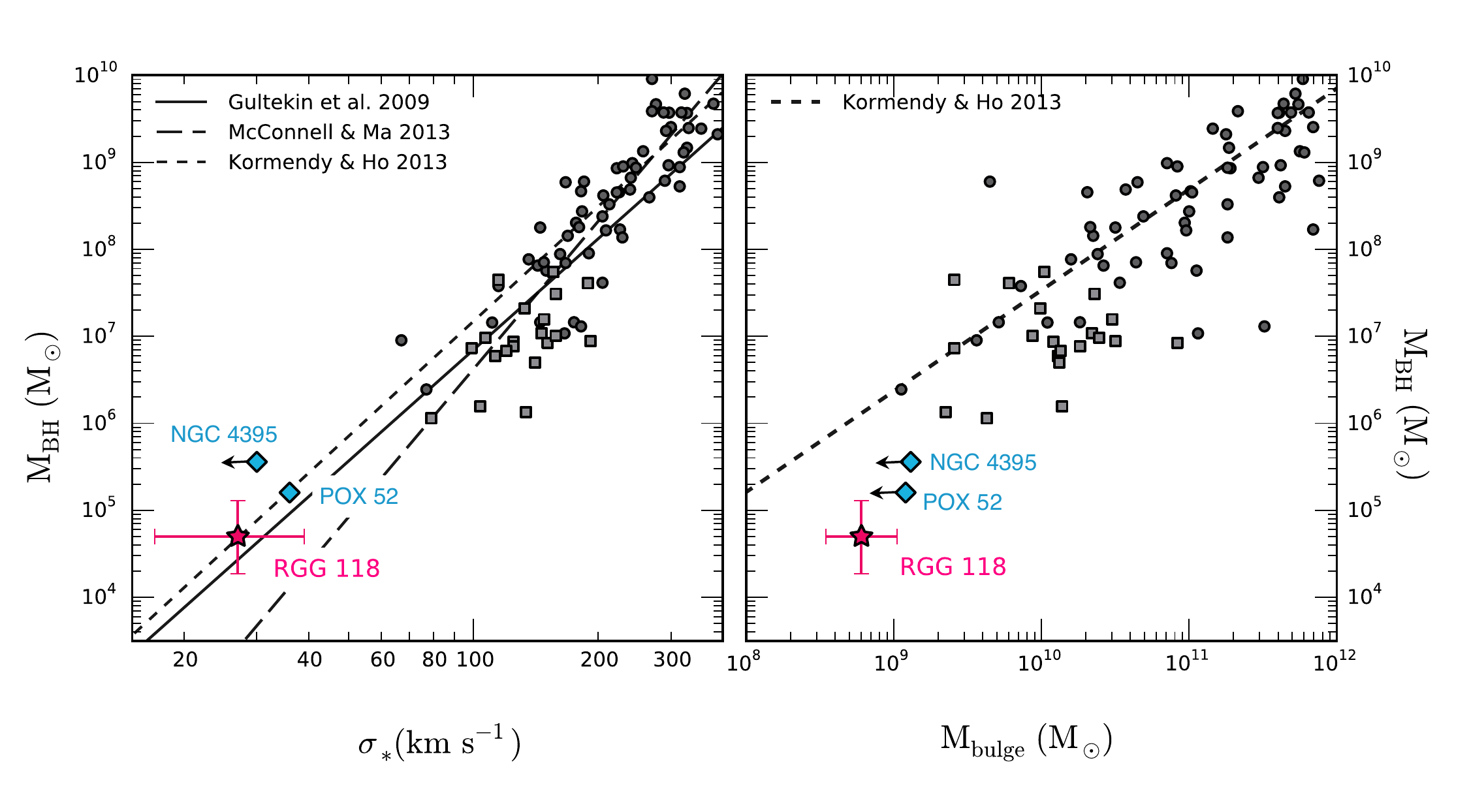}
\caption{{\it Both}: Black solid/dashed lines represent various determinations of scaling relations. Circles and squares represent systems having dynamically measured BH masses \citep{Kormendy:2013ve} with classical and pseudobulges, respectively. {\it Left}: $M_{\rm BH}-\sigma_{\ast}$ relation. RGG 118 is plotted as the pink star. Error bars on the mass account for the scatter in the correlations used to determine the BH mass; velocity dispersion errors include scatter in the relation between stellar and gas velocity dispersion \citep{2008AJ....136.1179B}. We also show the well-studied dwarf AGN POX~52 \citep{2004ApJ...607...90B, 2008ApJ...686..892T} and NGC~4395 \citep{2003ApJ...588L..13F} (turquoise diamonds). For NGC~4395, which is  bulgeless, the velocity dispersion is an upper limit and refers to the nuclear star cluster \citep{2003ApJ...588L..13F}. {\it Right}: Dashed black line gives the $M_{\rm BH}-M_{\rm bulge}$ relation. Total masses are plotted for POX~52 \citep{2008ApJ...686..892T}, and NGC~4395 \citep{2003ApJ...588L..13F}.}
\label{msigma}
\end{figure*}

We also find that RGG 118 falls on the extrapolation of the $M_{\rm BH}-\sigma_{\ast}$ relation to low BH masses, but falls an order of magnitude below the $M_{\rm BH}$--$M_{\rm bulge}$ relation (Figure~\ref{msigma}). We did not detect stellar absorption lines in the spectra of RGG 118, so instead use the gas velocity dispersion as a proxy for stellar velocity dispersion; the two have been shown to correlate, though the scatter in the relation is $\sim0.15$ dex \citep{2008AJ....136.1179B, 2011ApJ...739...28X}. 
Using the [NII] line, we measure a gas velocity dispersion of $\sigma_{\rm [NII]}=27 \kms \pm 5 \kms$, after correcting for instrumental dispersion.
This yields a stellar velocity dispersion estimate of $\sigma_{\ast}=27 ^{+12}_{-10} \kms$, after adding uncertainties in quadrature.

We compute a bulge mass of  $10^{8.8 \pm 0.2}M_{\odot}$ for RGG 118; this gives a $\rm M_{\rm BH}$-$\rm M_{\rm bulge}$ ratio of $\sim8\times10^{-5}$ and places RGG 118 roughly an order of magnitude below the $\rm M_{\rm BH}$-$\rm M_{\rm bulge}$ relation (Figure~\ref{msigma}). 
This is typical of galaxies with pseudobulges \citep{2008MNRAS.386.2242H}, which are more disk-like, flat, and rotationally dominated than classical galaxy bulges, and tend to have younger stellar populations \citep{Kormendy:2013ve}. We find the S\'{e}rsic index of the inner component to be n=$1.13\pm0.26$, which is consistent with a pseudobulge \citep{2004ARA&A..42..603K}. The high ellipticity of the bulge ($\epsilon = 0.63$) is also in accord with expectations for pseudobulges.

\section{Discussion}

Dwarf galaxies currently offer the best opportunity to understand BH seed formation and growth in the early universe. By itself, the BH in RGG 118 indicates that formation pathways must exist that produce BH seeds of its mass or less. Additionally, both the fraction of dwarf galaxies containing BHs and the slope/scatter of the low-mass end of $M_{\rm BH}-\sigma_{\ast}$ depend on the mechanism by which BH seeds form \citep{2008MNRAS.383.1079V, 2015ApJ...799...98M}. While searching for AGN in dwarf galaxies can produce only a lower limit on the BH occupation fraction for dwarf galaxies, it contributes toward constraining the low-mass end of $M_{\rm BH}-\sigma_{\ast}$. In particular, if BH seeds are generally massive (i.e. $\gtrsim10^{5}$ \msun) the slope of $M_{\rm BH}-\sigma_{\ast}$ is expected to flatten and the scatter to increase \citep{2009MNRAS.400.1911V}. We are in the process of measuring BH masses and velocity dispersions for additional targets identified by Reines et al. (2013). 

SDSS imaging reveals evidence for a pseudobulge in RGG 118. Pseudobulges do not seem to correlate with BH mass in the same way that classical bulges do. For a sample of BHs with $M_{\rm BH}<2\times10^{6}$ \msun, \cite{Greene:2008qy} found that most of the host galaxies were either compact systems or disk galaxies with pseudobulges. The BHs in these systems tended to be an order of magnitude less massive for a given bulge mass than those found in classical bulges. 
\cite{Kormendy:2013ve} argue that BHs in non-active galaxies with pseudobulges do not correlate at all with host galaxy properties, i.e. fall below both the $M_{\rm BH}-M_{\rm bulge}$ and $M_{\rm BH}-\sigma_{\ast}$ relations. However, they note that low-mass AGNs, which probe down to BH masses of $\sim10^{5}$ \msun~and $\sigma_{\ast}$ values of $\sim30\kms$, do seem to sit on the extrapolation of $M_{\rm BH}-\sigma_{\ast}$ while simultaneously falling below $M_{\rm BH}-M_{\rm bulge}$. This behavior is also seen for RGG 118. 
However, further work is needed to understand the nuclear structure of RGG 118. At a distance of $\sim100$ Mpc, the spatial resolution of the SDSS imaging ($\sim0.4\arcsec$) can only resolve structures with sizes of several hundred parsecs. Higher resolution imaging is needed to detect nuclear features such as disks and bars, which can bias estimates of bulge S\'{e}rsic index and mass and serve as strong indicators of a pseudobulge. \\

\vspace{0.2in}
The authors thank Belinda Wilkes and the Chandra X-ray Center for granting us Director's Discretionary Time. VFB is supported by the National Science Foundation Graduate Research Fellowship Program grant DGE 1256260. Support for AER was provided by NASA through Hubble Fellowship grant HST-HF2- 51347.001-A awarded by the Space Telescope Science Institute, which is operated by the Association of Universities for Research in Astronomy, Inc., for NASA, under contract NAS 5-26555.

\bibliographystyle{apj}

\end{document}